\magnification=1200
\hsize 5.5truein
\vsize 8.00truein
\hoffset 0.25truein
\voffset 0.25truein

   \baselineskip=14pt  
\overfullrule=0pt


\def\lap{\hbox{{\lower -2.5pt\hbox{$<$}}\hskip -8pt\raise -2.5pt\hbox{$\sim$}}}
\def\gap{\hbox{{\lower -2.5pt\hbox{$>$}}\hskip -8pt\raise -2.5pt\hbox{$\sim$}}}


\def\BB#1\par{
  \noindent\hangindent=20pt
  { #1 }
  \par
}
\def\Mesz{M\'esz\'aros\ }

\newcount\fignumber
\fignumber=1
\def\newfig{\global\advance\fignumber by 1}
\def\fignam#1#2{\xdef#1{\the\fignumber}\newfig}
\def\fignam#1#2{\xdef#1{\the\fignumber}\hfil\break [REMOVE FIG NAME #2:]\newfig}

%
\newcount\eqnumber
\eqnumber=1
%
\def\neweq{{\the\eqnumber}\global\advance\eqnumber by 1}
%
\def\eqnam#1#2{\xdef#1{\the\eqnumber}}

%
%
\def\lasteq{\advance\eqnumber by -1 {\the\eqnumber}\advance
     \eqnumber by 1}

\def\perscm{s$^{-1}$ cm$^{-2}$}

{}~~~~~~~~\ \ \
\bigskip
\bigskip
\bigskip
\bigskip
\bigskip
\centerline {\bf THE AVERAGE TEMPORAL AND SPECTRAL EVOLUTION}
\centerline{\bf OF GAMMA-RAY BURSTS}
\centerline {\bf E.~E.~Fenimore}
\centerline {{D436 Los Alamos National Laboratory, Los Alamos, NM 87545, USA}}
\centerline{email: efenimore@lanl.gov}
\centerline {Received..........;  accepted............}

\centerline {{\bf ABSTRACT}}

We have averaged  bright BATSE bursts to uncover the average
overall temporal and spectral evolution of gamma-ray bursts (GRBs).  We
align the temporal structure of each burst by setting its
duration to a standard
duration, which we call $T_{<{\rm Dur}>}$.
The observed average ``aligned $T_{<{\rm Dur}>}$'' profile for
32 bright bursts with intermediate
durations (16 to 40 s) has a sharp rise (within the first
20\% of $T_{<{\rm Dur}>}$)
and then a linear decay.  Exponentials and power laws do not fit this decay.
In particular, the power law seen in the x-ray afterglow ($\propto T^{-1.4}$)
is not observed during the bursts, implying that the x-ray afterglow is
not just an extension of the x-ray evolution seen during the gamma-ray
phase.
The average burst spectrum has a low energy slope of -1.03, a high
energy slope of -3.31, and a peak in the $\nu F_{\nu}$ distribution at
390 keV.
We determine the average
spectral evolution.  Remarkably, it is also a linear function, with
the peak of the $\nu F_{\nu}$ distribution given by 
 $\sim 680 - 600 (T/T_{<{\rm Dur}>})$
keV.
Since both the temporal profile and the peak energy are linear functions,
on average, the peak energy  is linearly proportional to the intensity.
This behavior is inconsistent with the external shock model.
Previously, trends have been reported for GRB evolution, but our
results are quantitative relationships that models should attempt to explain.
\hfill\break
{\it Subject Heading:} Gamma-Ray, Bursts
\hfill\break
\vfill\eject

\centerline{ 1.~INTRODUCTION}

Gamma-ray bursts (GRBs) are isotropically distributed yet are 
inhomogeneous implying  that they are much farther away than typical
galactic distance scales.  One explanation is that they are at
cosmological distances ($z \sim 1$).  At such distances, they require
$\sim 10^{51} (\Omega/4\pi)$ erg s$^{-1}$, where $\Omega/4\pi$ is the
fraction of the sky radiated into by the bursts.
There must exist a substantial
energy reservoir to power GRBs, and it is often suggested that
merging massive
objects are the ultimate source of energy (\Mesz \& Rees 1993; Piran,
Shemi, \& Narayan 1993; Rees \& \Mesz 1994).

GRBs are very diverse, with time histories ranging from as short as 50 ms to
longer than $10^3$ s.  The long bursts often have very complex temporal
structure with many subpeaks.  Two classes of theories have arisen
to explain how merging objects might produce the chaotic time history.
In one theory (\Mesz \& Rees, 1993), there is
a single release of energy ($\sim 10^{53}$ erg)
when two objects merge, which produces a relativistic shell.
The observation of very high energy photons, coupled with
the short time scales,
indicates that the relativistic shell has a bulk
Lorentz factor ($\Gamma$) of
$10^2$ to $10^3$.  
The shell expands for a rather long time (perhaps as long as $10^7$ s).
Eventually, the shell converts its bulk motion into gamma rays,
perhaps due to relativistic shocks caused by sweeping up the interstellar
medium (ISM).  As the shell emits, the shell keeps up
with the emitted photons,
so they all arrive at a detector over a short period of time. The duration 
of the event is set by the expansion of the shell. 
If $t_{\rm dur}$ is the length of time that the shells emits,
then the duration seen
at a detector is $T_{\rm dur} = t_{\rm dur}/(2\Gamma^2)$. In contrast, the 
complex temporal structure is due to inhomogeneities in the shell.

The alternative theory (Rees \& \Mesz 1994) is
that the merger takes place over a period of time
comparable to the observed duration of the burst.
Perhaps the massive object
is tidally disrupted, and the duration is set by the time it takes for
most pieces to merge.  Each subpeak in the GRB is the result of a separate 
explosive event at the central site.  We refer to this as a ``central
engine.''  A large Lorentz factor is still required, thus each subpeak
might produce a relativistic shell. 

In both theories, the chaotic time history involves some randomness in the 
process.  Neither involves a smooth process that produces a smooth envelope
of emission.  In this paper, we propose combining the time histories of
many GRBs to average out the randomness and uncover the mean
GRB envelope of emission. 

\centerline{ 2.~THE AVERAGE GRB ENVELOPE}

In the single relativistic shell scenario, the energy release is on 
a short time scale.
If the shell expands for a period of time before it converts its energy to 
gamma rays, the curvature of the shell imposes 
a ``FRED''-like shape to the time history.  (Here, ``FRED'' is a fast rise, 
exponential decay, although the decay might not actually be
exponential, it just falls faster than it rises.)
The duration of the FRED is directly related to the radius of the shell
at the time that it converts its energy (i.e., $2\Gamma^2cT_{\rm
dur}$).  This radius also could depend on the
density of the ISM ($\rho_{\rm ISM}$) around the bursts.  
If all
other things are equal, the observed burst duration from a single
relativistic
shell varies from burst to burst by some constant related
to $(2\Gamma^2)^{-1}$ and $\rho_{\rm ISM}$.
Thus, we scale the duration of each burst by a constant before
averaging them.  This can be viewed as an ``aligned $T_{<{\rm Dur}>}$''
average in contrast to the ``aligned peak'' averages, such as those used by
Mitrofanov, et al. (1996).  In the aligned peak average, each burst
contributes to the average by aligning the largest peak.  The time
scale of the peak is conserved as it contributes to the average.
In the aligned $T_{<{\rm Dur}>}$ average, each burst contributes to the
average by
aligning the midpoint of the burst, and the time scale of the burst is
adjusted to a standard duration which we call $T_{<{\rm Dur}>}$.

The Burst and Transient Source Experiment (BATSE)  catalog
provides durations called $T_{90}$ and $T_{50}$ (Meegan et al. 1996).
For example,
$T_{90}$ is the duration which contains 90\% of the counts.  It is
defined by finding the duration that excludes the first 5\% and last
5\% of the counts in the burst.  There is a similar definition for $T_{50}$.
Because of statistical variations in the background, it is impossible
to determine exactly when the burst starts and stops.  Thus,
insisting on observing some small fraction (like 5\%) is more reliable
than attempting to determine a duration which contains the whole event.
We estimate an average duration, $T_{<{\rm Dur}>}$, from $T_{90}$ and
$T_{50}$.  To first order,
$T_{<{\rm Dur}>}$ is $T_{90}/0.9$ or $T_{50}/0.5$.
By definition, the beginning
point for $T_{90}$ or $T_{50}$ must be at a point at which the count rate
is increasing. 
Thus, if we stretched each burst to a standard duration by scaling the
time by some multiple of $T_{90}$, there would be a coherent peak at the
first 5\% point and at the last 5\% point.
Rather, we define $T_{<{\rm Dur}>}$ to be a combination of
$T_{90}$ and $T_{50}$
to break up the coherency.  Specifically, we define
\eqnam{\TONE}{TONE}
$$
T_{<{\rm Dur}>} = {(T_{90} + T_{50})/2 \over 0.7}~~.
\eqno(\neweq)
$$
Although using the average tends to break up the coherency, some
coherency will still
remain and the average profile will tend to have a spike at the beginning.
If GRBs had box-car like time profiles,
then $T_{<{\rm Dur}>}$ would be equal
to $T_{90}/0.9 = T_{50}/0.5$. In fact, GRB profiles usually decay
after a fast rise such that $T_{90}/0.9 > T_{50}/0.5$. As a result,
$T_{<{\rm Dur}>}$ is between $T_{90}/0.9$ and  $T_{50}/0.5$.

We selected all 98 bursts from the BATSE 3B catalog that were longer than
1.5 s and brighter than 4 photons \perscm.
 To find the average profile of these
events, we first defined the time at the middle of each event, $T_{\rm
mid}$, to be the average 
of the $T_{90}$ and $T_{50}$ midpoints. (The BATSE catalog provides
these values, see Meegan et al. 1996.)
We used the 0.064 ms BATSE data summed from 25 to $\sim 1000$ keV and
rebinned
each burst from $T_{\rm mid}-T_{<{\rm Dur}>}$ to
$T_{\rm mid}+T_{<{\rm Dur}>}$ into
200 time samples each $T^{-1}_{<{\rm Dur}>}$ long.  This produces a scaled
time history from $T_{<{\rm Dur}>}/2$ before the start of the
$T_{<{\rm Dur}>}$ period
to $T_{<{\rm Dur}>}/2$ after the end of the $T_{<{\rm Dur}>}$ period.

The bursts have peak fluxes that range up to $10^2$ photons \perscm.
We normalize each burst so that they contribute to the average as if
they were all at the same distance.  This can be done in several ways.
 If
the ultimate energy source is either
neutron star-neutron star merger or neutron star-black hole merger,
the total
fluence is likely to be a standard candle.  Figure 1a gives the
average of the 98 bursts with the assumption that the fluence is a
standard candle; that is, we normalized the  total counts of a
burst to unity before it contributed to the average.
 Alternatively, it
is often assumed that the peak number flux is a standard candle.  This
is  required when applied to studies involving the triggering because
most detectors respond to peak counts.  We think that it is an
unlikely standard candle in scenarios involving merging objects.
Nevertheless, for completeness, in Figure 1b we normalized the 
peak number flux of a burst  to unity
before it contributed to the average.   
Although our intent was to normalize in order to scale all
bursts to the same
distance, we do not know if GRBs have standard candle fluence, standard
candle peak intensity, or if any characteristic is a standard candle.
However, if no characteristic is a standard candle, we would still normalize
the bursts such that no one burst would be allowed to dominate the average.
The most likely way to avoid a single burst dominating would be to normalize
by the total net counts.  Thus, in the absence of a known standard
candle, we would normalize as if the fluence was a standard candle.

Figure 1 shows bursts with $T_{90}$ durations from 1.5 s to 320
s. Figure  2 gives the aligned $T_{<{\rm Dur}>}$ average
time history using total fluence normalization for three ranges of
durations: 1.5 to 16, 16 to 40, and 40 to 320 s. Each group has about
33 bursts.
If one averaged Figures 2a, b, and c, the result would be Figure 1a.
  For
longer bursts, the trend is  to have a larger initial emission relative
to the rest of the burst.  The most reliable average is probably the
one that utilizes bursts with $T_{90}$'s between 16 and 40 s.  Bursts
with shorter durations often have only a few peaks, so the average
profile can contain large variations.
Also, short bursts do not have enough energy-resolved samples
(i.e., MER data)
to study the evolution of the spectra over the $T_{<{\rm Dur}>}$
range (see Fig.~3b below).
Bursts with longer durations tend to be complex, making it
difficult to
recognize the beginning and end of the bursts.  An additional
reason to emphasize the 16 to 40 s range is that it covers the
smallest dynamic range of durations (2.5 {\it vs.} 6.4 and 8.0 for the
1.5 to 16 and 40 to 320 s ranges, respectively).  This is a reflection
of the fact that the burst duration distribution peaks in the 16 to 40
s range so there are enough bursts to form an average even from a
small range of durations.  A small dynamic range means that the
difference in stretching from one burst to another is small
and, thus, the average should be more reliable.

For the reliable case of bursts with $T_{90}$ between 16 and 40 s,
the  average profile appears to peak about 20\% after the beginning of the
$T_{<{\rm Dur}>}$ period and then decay linearly to the
end of the $T_{<{\rm Dur}>}$ period.

\centerline{ 3. AVERAGE TEMPORAL AND SPECTRAL EVOLUTION}

In Figure 3 we used the BATSE MER data to investigate the average spectral
and temporal evolution. 
The BATSE MER data has 16 energy channels and 2.048 s time
resolution.
We used the 32 events with $T_{90}$ between 16 and 40 s because
we consider that average time history to be the most reliable (see above).
Figure 3a uses the same events as Figure 2b.  In Figure 3a, the poorer
time resolution of the MER data tends to wash out the large peak near
$T_{<{\rm Dur}>}$ = 0.2.  That peak is associated with the
coherent increase that occurs where the average  boundary of $T_{90}$ and
$T_{50}$ occurs.
The resulting average time history appears to rise to a peak and then fall
linearly.  We have fit a variety of temporal shapes to the decay portion
between 20\% after the beginning of the $T_{<{\rm Dur}>}$ period to the end
of the $T_{<{\rm Dur}>}$ period.
One cannot determine a strict goodness of fit
(e.g., $\chi^2$) because the uncertainties are not the counting statistics
(which, after all, are very good since we are adding together 32 bright
bursts).  Rather, the fluctuations are due to how the various peaks
of 32 bursts add together.  Presumably, if we had hundreds
of bursts, the curve would be very smooth.
However, we can use relative $\chi^2$ values to judge
relative goodness-of-fit.
We fit a linear function, an exponential function, and power law decays.
We particularly checked if a $T^{-1.4}$ power law would fit because that
type of decay is seen later during the x-ray afterglows (Piro et
al. 1997).
The linear fit was the best fit. The power law and exponential fit had
$\chi^2$ values that were 2.2 and 3,8, respectively, times larger and
they disagreed with the observations in a systematic way, failing to
agree with the observations at the ends of the time range.  A power
law with an index of -1.4 had a $\chi^2$ that was 6.4 times larger
than the linear fit.
The best linear function is:
$$
I = 5.56 {T \over T_{<{\rm Dur}>}}~~~~~~~~~~~~~~
{\rm if}~~ T < 0.18 T_{<{\rm Dur}>}
\eqno(\neweq a)
$$
$$
{}~~~~~~~=1.19 -1.06 {T \over T_{<{\rm Dur}>}}~~~~~~~
{\rm if}~~ T > 0.18 T_{<{\rm Dur}>}~~.
\eqno(\lasteq b)
$$

We calculated aligned $T_{<{\rm Dur}>}$ averages for each of the 16 MER
 energy channels. From these, 
 six spectra were formed, each  covering 15\% of the $T_{<{\rm Dur}>}$
range.
The first one started 5\% after the beginning of the $T_{<{\rm Dur}>}$
range, and the last one ended at 5\% before the end of the
$T_{<{\rm Dur}>}$ range.
For each of these spectra, we fit
the ``Band'' spectral shape (Band et al., 1993).
This shape consists of a low energy slope ($\alpha$), the peak of the
 $\nu F_{\nu}$ distribution ($E_p$),
and a high energy slope ($\beta$).  We first fit the Band shape to the
sum of all six spectra.  This yielded an overall average GRB spectra.  The
parameters are $\alpha = -1.03, E_p = 390$ keV, and $\beta=-3.13$.
Since this
is an average of 32 bright GRBs, it ought to provide a fairly good
average GRB spectra.
To investigate the average spectral evolution, we analyzed each of the
six spectra separately, fixing $\alpha$ to $-1.03$ and $\beta$ to $-3.13$.
Thus, the only free parameter is $E_p$.
 Figure 3b shows the resulting spectral evolution.
 It is
a remarkably straight line:
$$
E_p = 680 -600{T \over T_{<{\rm Dur}>}}~~~ {\rm keV}~~.  \eqno(\neweq)
$$

\bigskip
\centerline{ 4. DISCUSSION}

We have averaged bright BATSE bursts to uncover the average
overall temporal and spectral evolution of GRBs.
The expected profile from a single relativistic shell that expands in
a photon quiet phase and then becomes gamma-ray active (i.e., the external
shock model) should have an envelope with a  fast rise and
and a power law decay phase (i.e., $T^{-\alpha-1}$, where $\alpha$ is a
typical number spectral index for GRBs, say $\sim 1.5$). 
The peak of $\nu F_\nu$  should evolve as $T^{-1}$
(Fenimore, Madras, \& Nayakshin 1996; Fenimore \& Sumner 1997).
Figure 3a gives the observed average temporal profile for
32 bright bursts and
a power law decay is inconsistent with it.  Figure 3b gives the average
spectral evolution, and it is inconsistent with a $T^{-1}$ decay.
We conclude
that the average burst envelope does not support the external shock
model.
This disagreement adds to the arguments that the external shock model
cannot explain the observed GRB time histories
unless the shell has structure with angular scales much smaller than
$\Gamma^{-1}$
(see Fenimore, Madras,
\& Nayakshin 1996; Fenimore,
Ramirez, \& Sumner 1998).

The fact that the decay during the gamma-ray phase is inconsistent
with the decay during the x-ray afterglow (i.e., $T^{-1.4}$, see Costa
et al. 1997) argues that the x-ray afterglow is not just a
continuation of the evolution of the x-rays seen during the GRBs.

The observed decay is quite interesting.  Both the average temporal
evolution and the average spectral evolution are linear functions.
Thus, in the decay phase, the intensity is linearly proportional to the
peak of the $\nu F_{\nu}$ distributions.
The $\nu F_\nu$ evolution is effectively linear over the full duration
of the event.  In particular, the first point in Figure 3b occurs
during the rise in the time history.  This is consistent with
Norris et al. (1996)
which found that the the hardness often peaks before the peak in the
time history. Here, rather than just reporting a trend, 
we establish a quantitative relationship between
the  time history and spectral evolution.

Something in the physics determines the duration of events.
In the context of the external shock models, variations in $\Gamma$
from burst to burst
can easily cause the bursts observed in arrival time to have different
durations. We have argued that the GRB
phase is not a single relativistic shell, as in the external shock
model (Fenimore, Madras, \& Nayakshin 1996,
Fenimore, Ramirez, \& Sumner 1998,
and the results of this
paper). Rather, the GRB phase requires a central engine with energy
releases lasting up to $>100$ s
or the single relativistic shell must have structure much smaller than
$\Gamma^{-1}$.
  For example, this ``central engine'' might be
caused by  internal variations of $\Gamma$ 
in a relativistic wind.  It is not clear why these
variations should release energy such that, on average, the observed counts 
decreases
linearly and the peak of the energy spectrum also varies linearly.
The processes are random, and the peaks that are produced vary
substantially in intensity.  The result is that no one burst shows the
underlying linear relationships. When many bursts are added together,
the hidden pattern is revealed. 

 Although we have only shown the linear
relationships in the context of an average of many bursts, it seems
reasonable that this pattern exist in most of the bursts but is hidden by
the chaotic peak structure.  However, some bursts seem to show a
different pattern.  In an earlier report on this topic (Fenimore \&
Sumner 1997), we presented results like these shown in Figure 3 for fewer
bursts, where the linear trends were not as clear.  We also did an
aligned $T_{<{\rm Dur}>}$ average for six FRED-like bursts
(BATSE numbers 467, 543, 678, 2431, 2736, and 2994).
Only one of these bursts were included in Figure 3, the rest were too weak
to meet our criterion.
 In the case of
the FRED-like bursts, a
power law decay similar to that expected from an external shock model
was observed for the time history but not for the spectral evolution
(Fenimore \& Sumner 1997).

Two other quantitative relationships have been claimed for GRB
evolution.  Liang \& Kargatis (1996) suggested that $E_p$ varies as
$\exp^{-\Phi(T)/\Phi_0}$, where $\Phi(T)$ is the photon fluence up to
time $T$.  Given our equation (2), $\Phi(T)$ would scale as
$1.19T-0.53T^2$. Thus, we do not see the Liang \& Kargatis pattern in
the aligned $T_{<{\rm Dur}>}$ average profiles.  The other quantitative
pattern concerns how the duration of peaks varies with energy.  Peaks
are broader at lower energy, and Fenimore et al. (1995) found that
the GRB time structure scales as $E^{-0.45}$.  We have combined
equations (2) and (3) to characterize the width of the time structure as a
function of energy.  No pattern was evident; the aligned $T_{<{\rm Dur}>}$
linear relationship is not  the same as the aligned peak relationship.
Thus, if the physics is such that the aligned peak average is relevant, then
models must explain why the time structure scales as  $E^{-0.45}$.
If the physics is such that the aligned $T_{<{\rm Dur}>}$ average is relevant,
then models must explain why the time histories and spectral evolution
vary linearly with the duration of the events. Both relationships
should be consistent with the models since the aligned peaks
tend to be a feature of the largest peak, while the aligned $T_{<{\rm Dur}>}$
is a feature of the whole burst.

 We thank Jay Norris for providing the MER data.

\centerline{REFERENCES}

\BB
Band, D.~L., et al. 1993, ApJ, 413, 281

\BB
Costa, et al., 1997, Nature, 387, 783

\BB
{Fenimore, E.~E., Madras, C.~D., \& Nayakshin, S., 1996, ApJ 473, 998,
astro-ph/9607163}

\BB
Fenimore, E. E., \& Sumner, M. C., 1997,
All-Sky X-Ray Observations in the Next Decade, eds. Matsuoka, M.,
\& Kawai, N., in press,
astro-ph/9705052

\BB
Fenimore, E.~E., in't Zand, J.~J.~M., Norris, J.~P., Bonnell, J.~T.,
\& Nemiroff, R.~J., 1995, ApJ 448, L101, astro-ph/950407

\BB
Fenimore, E.~E.,
Ramirez, E., \& Sumner, M. C., Gamma-Ray Bursts: 4th Huntsville Symposium,
eds. Meegan, Preece, \& Koshut, in press, astro-ph/9712303.

\BB
Liang, E.~P., \& Kargatis, V.~E., 1996, Nature, 381, 49

\BB
Meegan, C.~A., et al. 1996, ApJS, 106, 65

\BB
\Mesz P., \& Rees, M. J., 1993, ApJ, 405, 278

\BB
Mitrofanov, I.~G., et al.,  1996, ApJ, 459, 570.

\BB
Norris, J.~P., et al. 1996, ApJ, 459, 393.

\BB
Piran, T., Shemi, A., \& Narayan, R. 1993, MNRAS, 236, 861

\BB
Piro, L., et al. 1997, A\&A, in press

\BB
Rees, M. J., \& \Mesz, P., 1994, ApJ, 430, L93


\centerline{\bf Figure Captions}

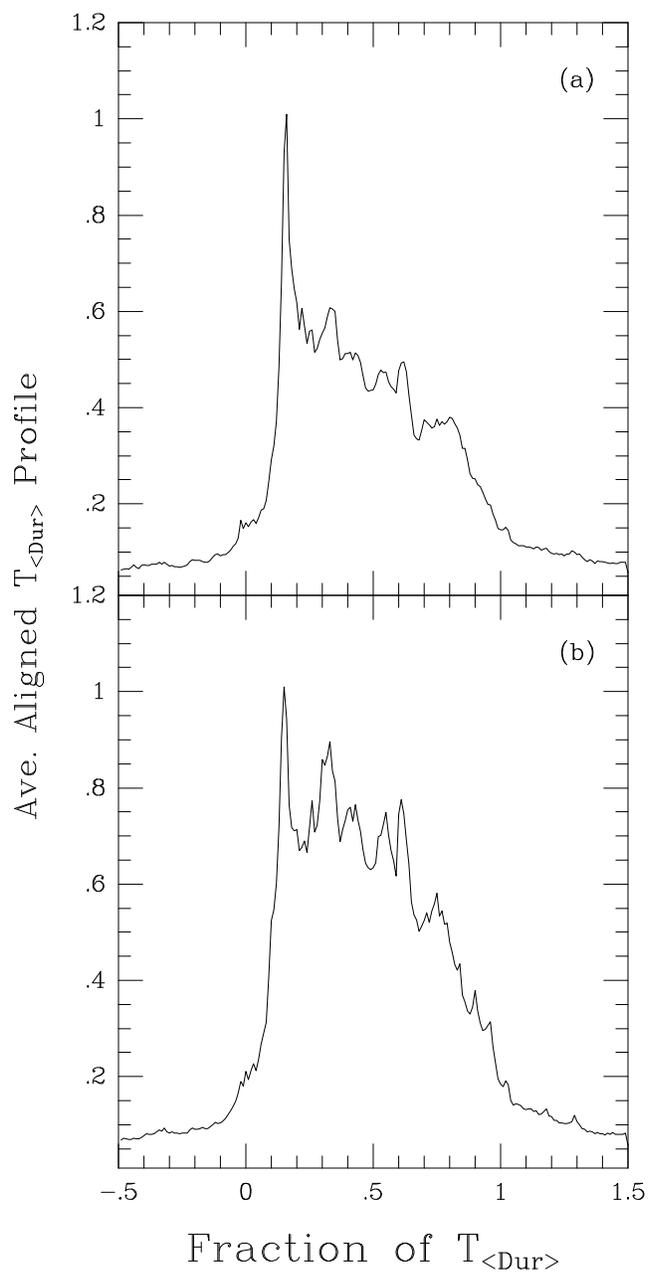
Figure 1: Average envelope of gamma-ray emission from 96 bright BATSE
bursts.
GRB models involving a single relativistic shell release energy
at the central site very quickly, and the duration of a GRB is set by
the evolution of the relativistic shell.  The emission from the shell
appears contracted in time due to relativistic
effects. To find the average time history envelope 
of emission in this case, we scale the observed duration of each GRB
(``$T_{<{\rm Dur}>}$'') to a standard duration.  This could  account for
different bulk Lorentz factors of the shells.
\hfill\break
(a) Each burst was normalized to the same distance by assuming a standard
candle total fluence.
\hfill\break
(b) Each burst was normalized to the same distance by assuming a
standard candle peak counts.

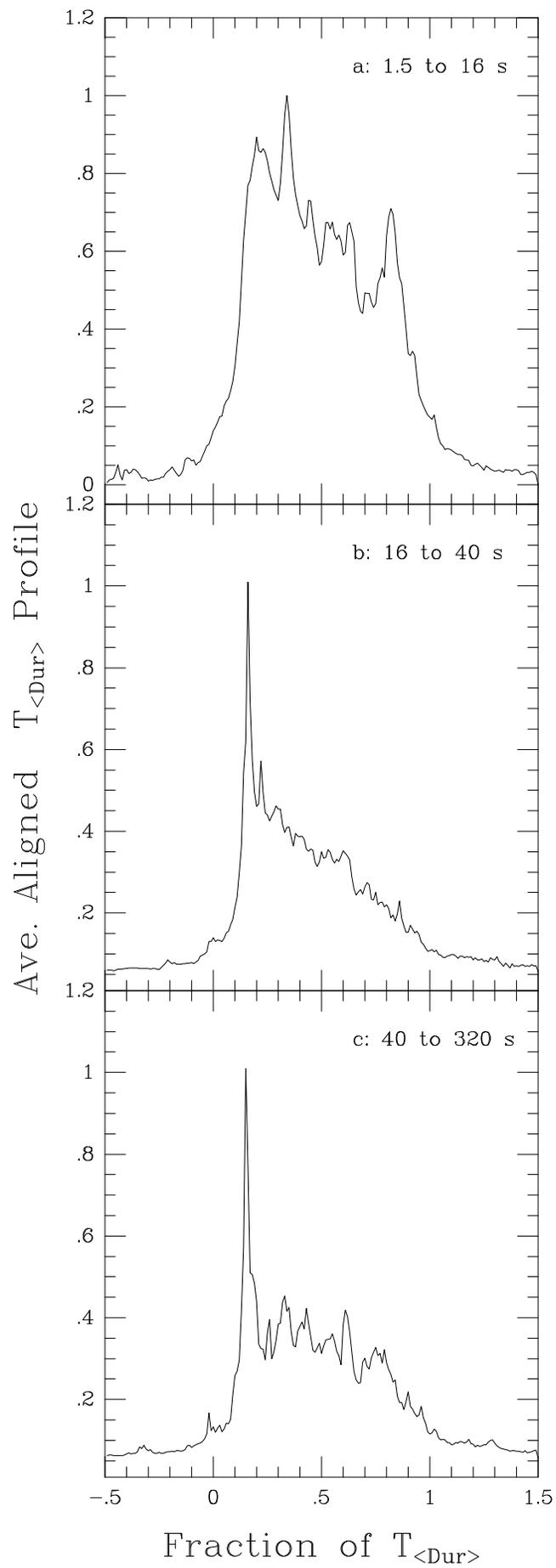
Figure 2: Average envelope of gamma-ray emission for three ranges of
observed durations.   Each burst was normalized to the same distance
by assuming a standard 
candle total fluence, and each duration was scaled to a
standard duration (``$T_{<{\rm Dur}>}$'').
 The average profile for bursts with duration between 16 and
40 s (b) is probably the most reliable ``aligned $T_{<{\rm Dur}>}$''
average profile.

Figure 3: The average temporal and spectral evolution of bright events with
intermediate durations ($T_{90}$ between 16 and 40 s) based on the
BATSE MER data.
\hfill\break
(a) The average time history.  The decay phase starting 20\% after the
beginning of the $T_{<{\rm Dur}>}$ period is
inconsistent with exponential decays and
power law decays (including the $T^{-1.4}$ decay seen  in the
x-ray afterglow phase).  Instead, the decay is consistent with a linear
slope.
\hfill\break
(b) The average spectral evolution.  The spectral evolution is found by
fixing the low energy and high energy slopes at the average for the
bursts and allowing only the peak of the $\nu F_\nu$ to vary.  
The peak energy is also a linear function.  
Thus, on average, the intensity is a
linear function of the peak of the $\nu F_{\nu}$ distribution.  This pattern
is inconsistent with that expected from the external shock model.

\vfill\eject
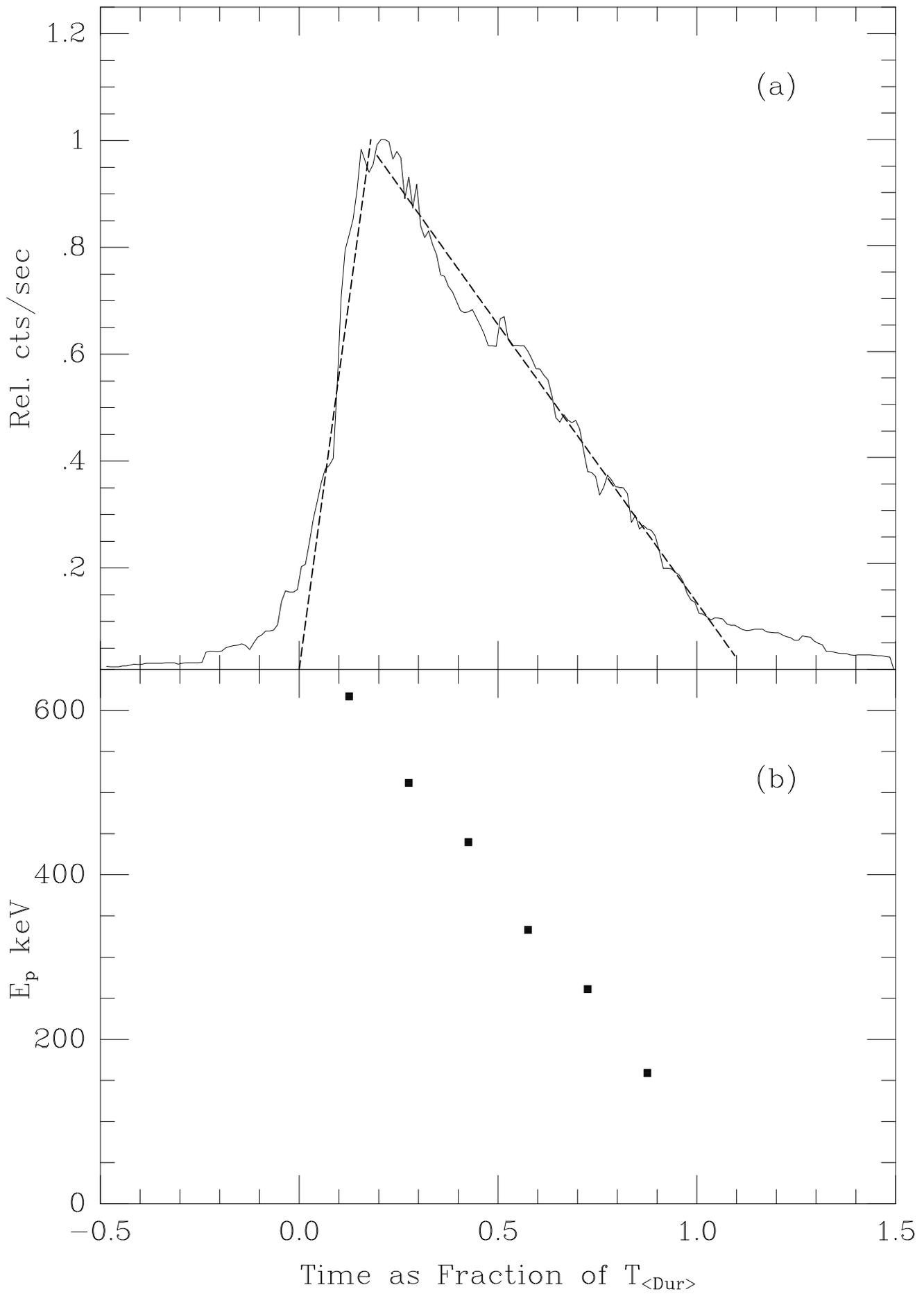
\bye